\documentclass[trackchanges,twocolumn]{aastex7}

\usepackage{enumitem}
\usepackage{subcaption}

\begin{document}

\title{Brown Dwarf Formation Through Gravitational Collapse: Insights From 3D Numerical Simulations}

\author[orcid=0009-0004-2719-1107,sname='Ahmad']{Adnan Ali Ahmad}
\affiliation{Univ Lyon, Ens de Lyon, Univ Lyon 1, CNRS, Centre de Recherche Astrophysique de Lyon UMR5574, 69007, Lyon, France}
\email[show]{adnan-ali.ahmad@cnrs.fr}

\author[orcid=0000-0003-2407-1025, sname='Commerçon']{Benoît Commerçon} 
\affiliation{Univ Lyon, Ens de Lyon, Univ Lyon 1, CNRS, Centre de Recherche Astrophysique de Lyon UMR5574, 69007, Lyon, France}
\email{fakeemail2@google.com}

\author[orcid=0000-0002-8342-9149,sname=Chabrier]{Gilles Chabrier}
\affiliation{Ecole normale supérieure de Lyon, CRAL, Université de Lyon, UMR CNRS 5574, F-69364 Lyon Cedex 07, France}
\affiliation{School of Physics, University of Exeter, Exeter, EX4 4QL, UK}
\email{chabrier@ens-lyon.fr}

\author[orcid=0009-0005-5957-9429,sname=Borderies]{Antonin Borderies}
\affiliation{Univ Lyon, Ens de Lyon, Univ Lyon 1, CNRS, Centre de Recherche Astrophysique de Lyon UMR5574, 69007, Lyon, France}
\email{chabrier@ens-lyon.fr}

\begin{abstract}

The formation mechanism of Brown Dwarfs (BDs), whether akin to stars or ejected planetary-mass objects, remains debated. We present the first 3D radiation-MHD simulations of magnetized, turbulent, gravitationally unstable low-mass cores ($0.05-0.1\ \mathrm{M_{\odot}}$) collapsing into proto-BDs. Using the {\ttfamily RAMSES} code with adaptive mesh refinement, we model the full dynamical range ($10^{5}~-10^{22}\ \mathrm{cm^{-3}}$), including radiative transfer (flux limited diffusion) and non-ideal MHD (ambipolar diffusion). Our simulations self-consistently follow the isothermal collapse, first hydrostatic core formation, H$_{2}$ dissociation, and BD birth. The resulting BDs have initial radii $\approx 0.75\ \mathrm{R_{\odot}}$ and masses $\approx 0.8\ \mathrm{M_{Jup}}$, growing via accretion as we follow the early evolution of the object. Crucially, we find that BDs may form similarly to low-mass stars but with a prolonged first-core phase, supporting a star-like formation scenario.

\end{abstract}

\section{Introduction} 
\textit{}
%Main Text – no more than 3500 words (not including acknowledgments, appendices or other supplementary material)
%Figures and Tables – no more than 5 combined figures (each limited to 9 panels) and tables, e.g. 3 figures and 2 tables.
Brown Dwarfs (BDs) are the lowest mass star-like objects that are able to form in the Universe, with a mass range that precludes them from sustaining hydrogen fusion in their core, although deuterium burning may still occur. First proposed by \cite{kumar_1963, hayashi_1963}, they were later discovered by \cite{rebolo_1995, nakajima_1995, oppenheimer_1995}. These enigmatic objects are currently the subject of debate in the literature, as a consensus is yet to be achieved regarding the dominant formation mechanism behind them, and hence, how they shape the lower end of the initial mass function (IMF). 
\\
\indent Three formation scenarios dominate the discourse; these include the fragmentation of a gravitationally unstable circumstellar disk \citep{stametellos_2007, stametellos_2008}, stunted growth due to competition from other protostars or ejection from their host environments \citep{reipruth_2001, bate_2002, bonnell_2006, bonnell_2008, basu_2012, reirpurth_2015, bate_2019, coleman_2025}, and the star-like scenario where Brown Dwarfs form as a scaled-down version of star formation, where gravo-turbulent processes create the necessary density peaks in the interstellar medium \citep{padoan_2002, padoan_2004, hennebelle_2008c, chabrier_2010, hopkins_2012, chabrier_2014, haugbolle_2018, vazquez_2019, dhandha_2024}. From an observational point of view, a number of advances, particularly with Herschel, ALMA, JWST and Euclid, currently point towards the star-like scenario as the most likely explanation for current BD distribution (see the reviews by \citealp{chabrier_2014, palau_2024}). Of noticeable importance is the first observation of an isolated proto-BD core of $\sim 0.02-0.03~\mathrm{M_{\odot}}$ and mean density $\sim 7.5 \times 10^6~\mathrm{cm^{-3}}$ in the star-forming cluster Ophiucus by \cite{andre_2012}. In addition, a recent study by \cite{redaelli_2025} has reported a density peak of $\sim 10^{7}\ \mathrm{cm^{-3}}$ in Corona Australis 151, showing that it is indeed possible to form the Jeans-unstable structures required to form a BD through an isolated collapse. However, it is important to note that other scenarios have been suggested for the lowest mass BDs. This is apparent in the reported excess of very low-mass objects compared with the standard IMF \citep{barrado_2001, Moraux_2003, oasa_2008, lodieu_2011, ramirez_2012, bouy_2015, drass_2016, miret_2022, miret_2023, Kirkpatrick_2024, luhman_2025, defurio_2025, Muzic_2025}, which point at the existence of a planet-like formation scenario. Recently, JWST observations by \cite{Haworth_2025} have also shown a dark tail behind a BD, possibly hinting at a recent ejection from its nascent environment.
\\
\indent Under the star-like formation scenario, a BD would form by following the classical \cite{larson1969} sequence, in which the collapse proceeds in two steps; forming a first core in hydrostatic equilibrium after radiative cooling becomes inefficient, and a second core when H$_2$ dissociation is completed. However, theoretical works having studied the collapse of a pre-stellar core in the BD mass regime are sparse. Works having done so assuming 1D spherical symmetry report that BD formation through isolated collapse requires unusually strong turbulent compression \citep{lomax_2016, stamer_2018, stamer_2019}, however, they find that once the collapse proceeds, it resembles the classical \cite{larson1969} sequence, albeit with a significantly longer first core lifetime\footnote{See also the lowest mass run presented in \cite{vaytet_2017}.}. \cite{tomida_2010b} performed 3D RHD calculations of a $0.1\ \mathrm{M_{\odot}}$ core and found that the first core has a significantly extended lifetime owing to lower mass accretion rates. \cite{machida_2009} performed 3D MHD calculations describing the collapse of a $0.22\ \mathrm{M_{\odot}}$ cloud using a barotropic equation of state (EOS) and Ohmic dissipation, and they reported that magnetic braking increases the mass accretion rate on the first core, allowing it to trigger a second collapse sooner. Although they report the existence of a protostellar outflow as well that of a circumstellar disk, they use a sink particle to describe the nascent BD.
\\
\indent Notably, there is no 3D simulation having resolved a nascent BD without the use of any sub-grid modeling (such as a sink particle). In the present study, we present the first 3D calculations having described the birth of a BD through a gravitational collapse fully self-consistently (i.e., without the use of a sink particle), while accounting for non-ideal magnetic fields and radiative transfer.

\section{Collapse with {\ttfamily RAMSES}} \label{sec:style}

Using the {\ttfamily RAMSES} adaptive mesh refinement (AMR) code \citep{teyssier_2002}, we model the collapse of turbulent and gravitationally unstable dense cores of masses $0.1\ \mathrm{M_\odot}$ and $0.05\ \mathrm{M_\odot}$, while accounting for non-ideal MHD with ambipolar diffusion \citep{teyssier_2006, Fromang_2006, masson_2012} and radiative transfer under the gray flux-limited diffusion approximation \citep{commercon_2011, commercon_2014, gonzalez_2015}. Our use of the realistic equation of state of \cite{saumon_1995} allows us to accurately model the phase transitions of the fluid during the collapse, as well as accurately describing the internal structure of the nascent BD. We use the opacity table of \cite{vaytet_2013} for our radiative transfer.

%% The "ht!" tells LaTeX to put the figure "here" first, at the "top" next
%% and to override the normal way of calculating a float position.
%% The asterisk after "figure" tells the compiler to span multiple columns
%% if a two column style is selected.
% \begin{figure*}[ht!]
% \plotone{AuthorChargeInfographic.png}
% \caption{The AAS journals are operated as a nonprofit venture, and author charges fairly recapture costs for the services provided in the publishing process. The chart above breaks down the services that author charges go toward. The AAS Journals' Business Model is outlined in a \href{https://aas.org/posts/news/2023/08/aas-open-access-publishing-model-open-transparent-and-fair}{2023 post}.
% \label{fig:general}}
% \end{figure*}

\subsection{Initial conditions}
We model the collapse of an isolated homogeneous sphere of temperature $T_{0}$=10 K and mass $M_{0}$, that is gravitationally unstable with a thermal-to-gravitational energy ratio of
\begin{equation}
    %\alpha = \frac{5R_{0}k_{\mathrm{B}}T_{0}}{2GM_{0}\mu m_{\mathrm{H}}},
    \alpha = \frac{5R_{0}}{2GM_{0}}c_{s_{0}}^{2},
\end{equation}
where $R_{0}$ is the radius of the dense core, $G$ the gravitational constant, and $c_{s_{0}}=1.9\times10^{4}\ \mathrm{cm\ s^{-1}}$ the initial sound speed. The sphere is threaded along the $z$ axis with a uniform magnetic field whose strength is parametrized by the mass-to-flux ratio \citep{mouschovias_1976}
\begin{equation}
    \mu_{\mathrm{m}} = \frac{M/\phi_{\mathrm{B}}}{(M/\phi_{\mathrm{B}})_{\mathrm{crit}}},
\end{equation}
\\
where $\phi_{\mathrm{B}} = \pi r^{2}B$ is the magnetic flux. We introduce an initially turbulent velocity vector field randomly sampled from a 3D Gaussian distribution with a Kolmogorov power spectrum ($P(k)\propto k^{-11/3}$). It is parametrized by the turbulent Mach number $\mathcal{M}$\footnote{All five runs use the same initial seed for the turbulent velocity field.}. We do not consider solid body rotation. We present in this letter five runs with the initial conditions presented in Table \ref{tab:IC}.

\begin{deluxetable*}{cccccc}
\tablecaption{Initial conditions of the 5 runs presented in this letter, and the resolution at the finest AMR level.\label{tab:IC}}
\tablehead{
\colhead{Run Label} & 
\colhead{$M_{0}$} & 
\colhead{$\mathcal{M}$} & 
\colhead{$\mu_{m}$} & 
\colhead{$\alpha$} & 
\colhead{$\Delta x_{\mathrm{min}}$} \\
\colhead{} & 
\colhead{($M_{\odot}$)} & 
\colhead{} & 
\colhead{} & 
\colhead{} & 
\colhead{($R_{\odot}$)}
}
\startdata
R1 & 0.05 & 0.4 & 4 & 0.25 & $1.2\times10^{-2}$ \\
R2 & 0.1 & 0.4 & 4 & 0.25 & $5.0\times10^{-2}$ \\
R3 & 0.1 & 0.1 & 4 & 0.25 & $2.5\times10^{-2}$ \\
R4 & 0.1 & 0.4 & 8 & 0.25 & $2.5\times10^{-2}$ \\
R5 & 0.1 & 0.4 & 4 & 0.5 & $5.0\times10^{-2}$ \\
\enddata
\end{deluxetable*}

% \begin{table}[h!]
%     \centering
%     \begin{tabular}{| c | c | c | c | c | c |}
%         \hline
%           Run label & $M_{0}$ [$\mathrm{M_{\odot}}$] & $\mathcal{M}$ & $\mu_{m}$ & \add{$\alpha$} & $\Delta x_{\mathrm{min}}\ \mathrm{[R_{\odot}]}$\\
%          \hline
%          R1 & 0.05 & 0.4 & 4 & \add{0.25} & $1.2\times 10^{-2}$\\
%          \hline
%          R2 & 0.1 & 0.4 & 4 & \add{0.25} & $5\times 10^{-2}$\\
%          \hline
%          R3 & 0.1 & 0.1 & 4 & \add{0.25} & $2.5\times 10^{-2}$\\
%          \hline
%          R4 & 0.1 & 0.4 & 8 & \add{0.25} & $2.5\times 10^{-2}$\\
%          \hline
%          R5 & 0.1 & 0.4 & 4 & \add{0.5} & $5\times 10^{-2}$\\
%          \hline
%     \end{tabular}
%     \caption{Initial conditions of the \add{5} runs presented in this letter, and the resolution at the finest AMR level.}
%     \label{tab:IC}
% \end{table}
We refine the grid according to the Jeans length $\lambda_{\mathrm{J}}$, where we enforce 20 cells per $\lambda_{\mathrm{J}}$\footnote{$\lambda_{\mathrm{J}}$ is computed at a fixed temperature of 100 K when $T>100$~K (see \citealp{ahmad_2023}), yielding $>$ 200 cells per true $\lambda_{\mathrm{J}}$ throughout the second collapse.}. The coarse grid begins at $\ell_{\mathrm{min}}=6$ and we allow up to 16 levels of mesh refinement, yielding a resolution at the finest level that is displayed in the final column of Table \ref{tab:IC}. Aggressive refinement criteria are a necessity to accurately capture all relevant physical processes during the collapse \citep{commercon_2008, vaytet_2017, vaytet_2018, federrath_2011, federrath_2014}.

\subsection{First core lifetime and second collapse}
    
\begin{figure}[h]
\centering
\includegraphics[scale=.32]{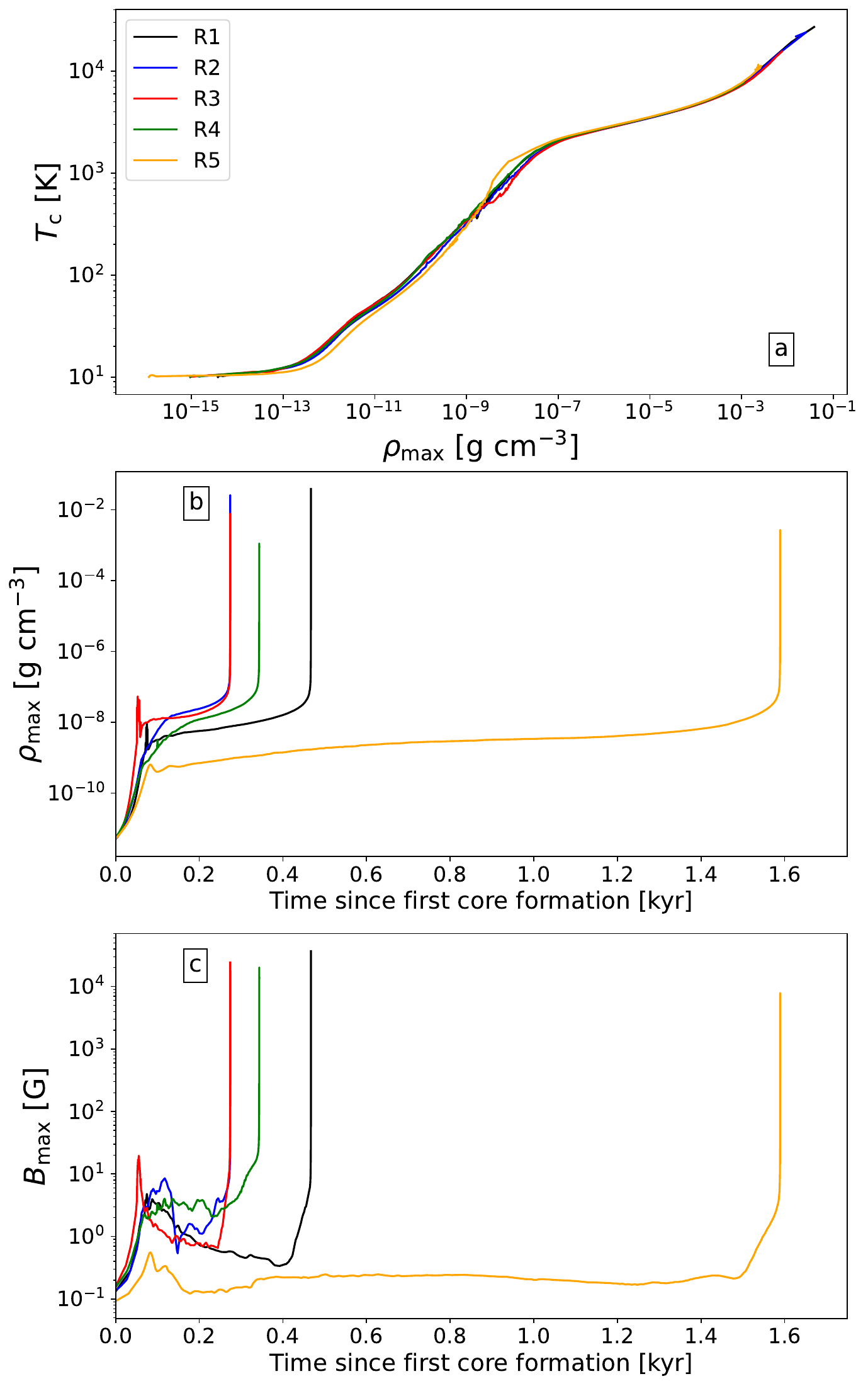} %pour fixer l'échelle.
\caption{A timeline of the evolution of collapse. Panel (a) displays the temperature-density evolution of the densest cell in our runs. Panels (b,c) display the maximum density and magnetic field strength as a function of time since first core formation, respectively.}
\label{fig:centralprops}
\end{figure}

The collapse proceeds following the classical \cite{larson1969} sequence, as seen in in Fig. \ref{fig:centralprops}a, which shows the temperature-density evolution of the densest cell in our runs. All five runs formed a BD, however the lifetime of the first core varies, which can be seen in Fig. \ref{fig:centralprops}b,c. These display the maximum density and maximum magnetic field strength as a function of time after first core formation. Steep rises in maximum density correspond to the onset of the second collapse. Run R1, having the lowest mass, exhibits the longest first core lifetime when compared to runs R2-4 as expected, in line with the results of \cite{tomida_2010b, stamer_2018}. Runs R2-4 all have the same mass, and hence the longer first core lifetime in run R4 may be explained by the weaker magnetic field, which reduces the efficiency of magnetic braking and hence leads to lower mass accretion rates on the first core when compared to run R2-3. The first core lifetimes in runs R2-3 are the same, showing that the additional angular momentum in run R3 is not sufficient to extend the first core lifetime because of magnetic braking. Finally, run R5 exhibits the longest first core lifetime, reaching $\sim 1.6$~kyr as a result of increased thermal pressure support in the dense cloud core.
\\
The maximum magnetic field strength in the first core is of the order of $\sim 1$ G owing to ambipolar diffusion\footnote{The field strength at the densest cell is of the order of $\sim 0.5$ G.}. The magnetic field surrounding the first core drives a bipolar low velocity outflow (see Appendix \ref{appendix:outflow}). Once the second collapse proceeds, the ideal MHD approximation is recovered and a maximum field strength of $\sim 10^{4}$ G is implanted in the proto-BD. These results are similar to the $1\ \mathrm{M_{\odot}}$ calculations presented in \cite{ahmad_2025}, and consistent with observations of cool BDs (\citealp{hallinan_2008, kao_2018}). However, whether this implanted fossil field is the one that is measured in observations remains unclear.

\subsection{The nascent Brown Dwarf}

\begin{figure*}%[h]
\centering
\includegraphics[scale=.5]{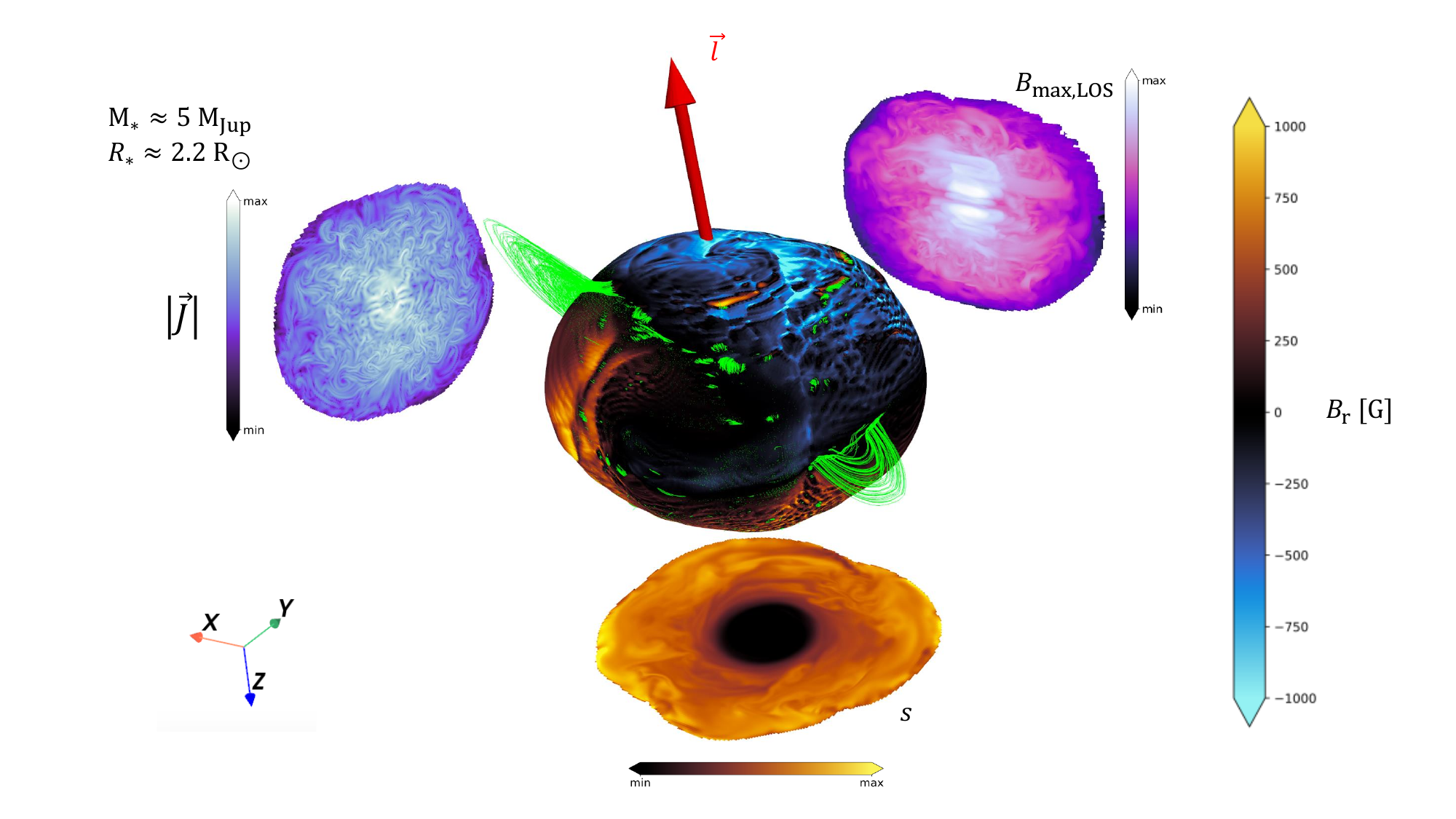} %pour fixer l'échelle.
\caption{A 3D rendering showing the surface of the BD in R1 when it is two weeks old. The color scale displays the radial magnetic field strength on the surface. The green streamlines display closed magnetic loops, meaning field lines that loop from magnetic north to magnetic south on the surface. The red arrow displays the angular momentum axis of the BD. The left and bottom image panels are cross-sectional slices of the interior of the BD, displaying the magnitude of current density (left) and specific gas entropy (bottom). The image panel on the right displays the maximum magnetic field strength along the line of sight (y axis). The mass and radius of the BD are displayed in the top left corner.}
\label{fig:3D}
\end{figure*}

In Fig. \ref{fig:3D}, we display a 3D rendering of the BD \footnote{We define the BD as all the gas whose fraction of dissociated H$_2$ exceeds 95\%.}in run R1 approximately two weeks after its birth, onto whose surface we encode the radial magnetic field strength, which reaches values of $\sim 1\ ~\mathrm{kG}$. The field is also dipolar, with a magnetic north and south pole which coincidentally is anti-aligned with the angular momentum axis of the BD (red arrow in Fig. \ref{fig:3D}). The left image panel displays a slice showing the current density within the BD, which exhibits how turbulent the field is in the interior. This is the result of accretion driven turbulence \citep{klessen_hennebelle}, which was previously reported in low-mass protostars \citep{bhandare_2020, ahmad_2023}. The strong turbulent driving causes significant entropy mixing, as can be seen in the bottom image panel of Fig. \ref{fig:3D}, where turbulent cells mechanically transport energy, and hence the entropy content of the fluid is moved. Figure \ref{fig:3D} also displays the maximum field strength throughout the interior (right image panel), which exhibits a turbulent and mainly toroidal field threaded with a significant poloidal component. The ratio of azimuthal to poloidal components vary and is of the order of $\sim 3$ in the interior. The green streamlines are closed magnetic loops, which show magnetic field lines looping from magnetic north to south, akin to solar current sheets. Although the BD's implanted magnetic field is dipolar, we do not see a large scale dipole with parallel loops, as many magnetic field lines remain open at the surface. We report that this magnetic field changes over time, and it is currently the subject of a study whose results will be presented in a follow-up paper.
\\

\subsection{Accretion and internal structure}

\begin{figure}%[h!]
\centering
\includegraphics[scale=.32]{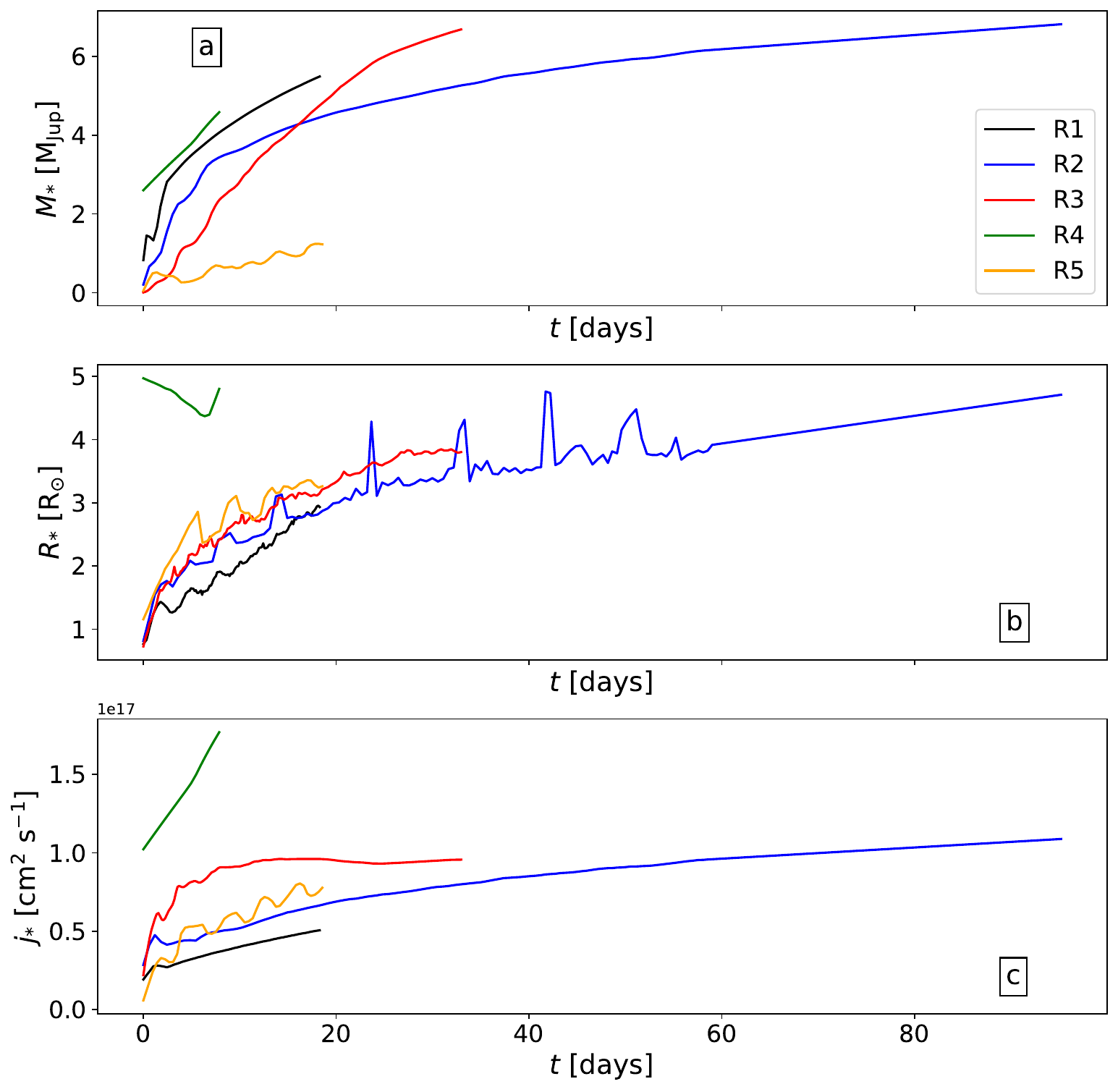} %pour fixer l'échelle.
\caption{Mass (a), radius (b), and specific angular momentum of the BD in each of our runs, displayed as a function of time since BD birth.}
\label{fig:BDprops}
\end{figure}

\begin{figure*}%[h!]
\centering
\includegraphics[scale=.2]{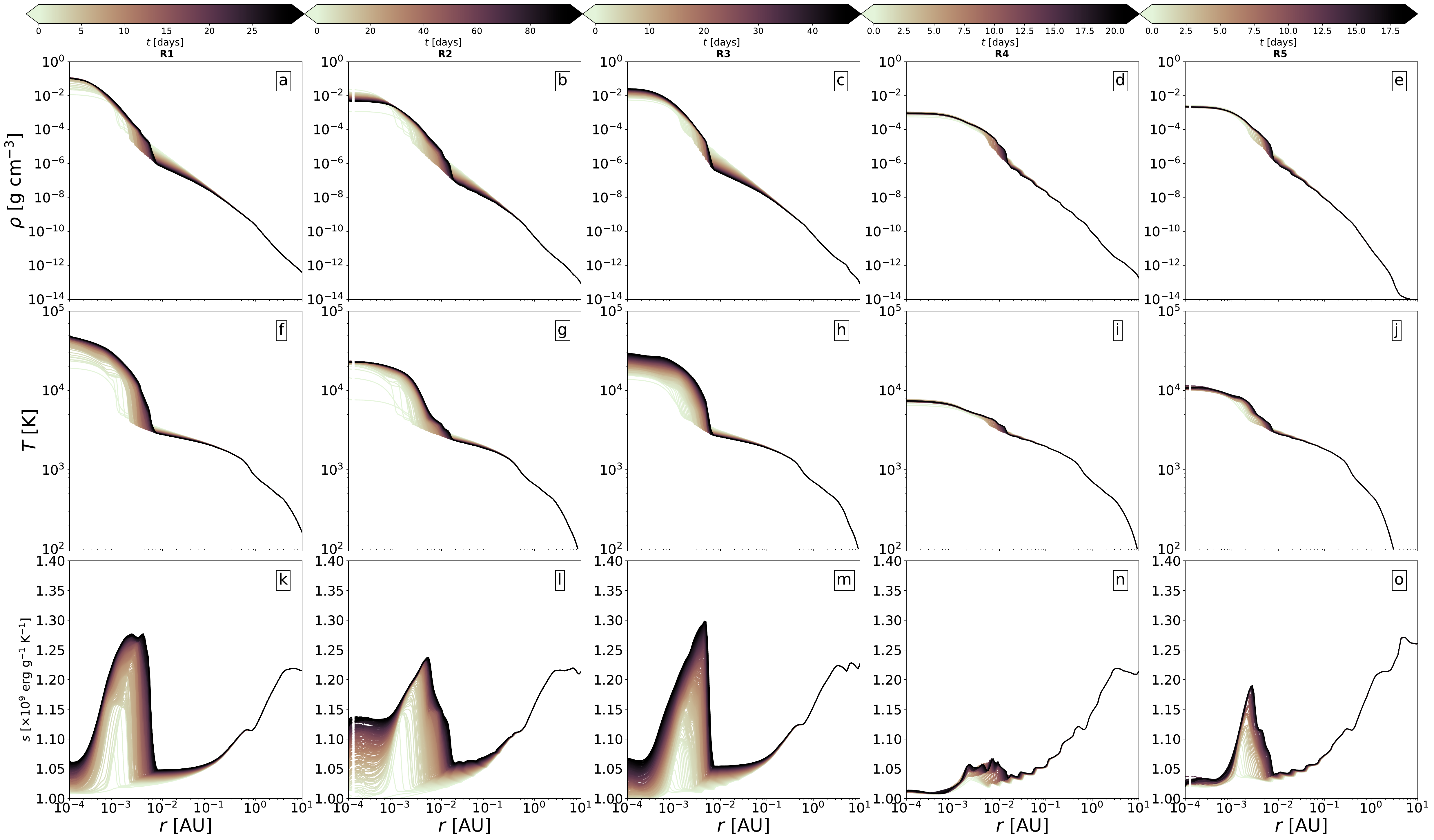} %pour fixer l'échelle.
\caption{Evolution of the density (first row, a-e), temperature (second row, f-j) and specific entropy (third row, k-o), averaged in radial bins and displayed as a function of radius at different times, where $t=0$ marks the birth of the BD. Each column corresponds to a different run.}
\label{fig:profiles}
\end{figure*}

After its formation, the BD accretes a significant amount of material from the remnants of the first core. This results in a significant mass increase immediately following birth, as seen in Fig. \ref{fig:BDprops}a. The mass increase also results in a swelling of the BD (Fig. \ref{fig:BDprops}b), as well as an increase in the angular momentum content of the BD\footnote{Such a rapid accretion of angular momentum by the BD will lead to a rotational breakup and subsequent disk formation, similarly to that seen in \cite{machida_2011b, ahmad_2024}. See Appendix \ref{appendix:diskform}.}. We also notice oscillations of the BD radius in runs R2-5, which was previously seen in the high resolution hydrodynamical study of \cite{ahmad_2023}.
\\
We also notice that runs R1-3 and R5 follow a different evolutionary trend to R4, which due to its weaker magnetic field, contains far greater amounts of angular momentum. Run R5 also exhibits a lower mass and slower mass increase when compared to runs R1-4, due to the increased thermal energy present in the gas, which further swells the BD.
\\
Despite the similar evolutionary trends, the interior structure of the BD shows some differences. In Fig. \ref{fig:profiles}, we display the average density (a-d), temperature (e-h), and specific entropy (i-l) for each run as a function of radius at different times. Steep gradients in these three quantities corresponds to the location of the accretion shock, which is the BD's boundary. We see that each run displays differing density-temperature profiles. The entropy profile for runs R1 and R3 show that the BD is radiatively stable \citep{schwarzschild_1906, ledoux_1947}, although runs R2, R4, and R5 seem to show some convective instability in their outer regions, as their entropy gradient is negative there. The interior structure of run R4 displays results that are much closer to hydrodynamical runs \citep{machida_2011, Bate_2011, ahmad_2024}, as the density and temperature profiles are much flatter owing to increased angular rotation rates when compared to runs R1 and R3.

\section{Conclusion} \label{sec:floats}

We have carried out the first 3D simulations that self-consistently describe a Brown Dwarf born out of a gravitational collapse of an isolated dense core, while accounting for non-ideal magnetic fields with ambipolar diffusion and radiative transfer under the gray FLD approximation. The collapse follows the isothermal phase, the first adiabatic phase, the second collapse triggered by H$_2$ dissociation, and the birth of the Brown Dwarf.
\\
In the midst of the debate regarding their dominant birth mechanism, these results provide insights into brown dwarfs born out of a gravitational collapse. Our findings can be summarized as follows:
\begin{enumerate}[label=\roman*]
    \item The collapse of very low-mass dense cloud cores follows the classical \cite{larson1969} sequence, albeit the first core lifetime is longer owing to its reduced mass accretion rate. The initial mass content of the dense core, its thermal energy content, as well as the magnetic field strength are the dominant factors in determining the first core lifetime.
    \item Nascent Brown Dwarfs have typical masses of $\approx~0.8~\mathrm{M_{Jup}}$ and radii $\approx~0.75~\mathrm{R_{\odot}}$. They experience high mass accretion rates and grow significantly after birth, and their angular momentum content also shows a significant increase.
    \item The magnetic field strength implanted in the Brown Dwarf at birth is of the order of $\sim 1~\mathrm{kG}$ at the surface. The field is primarily dipolar, with current sheets protruding from the surface. However, the surrounding magnetized gas disrupts the dipole’s loops from closing, preventing the formation of a large-scale magnetic arcade by stretching or opening the field lines. In the interior, the magnetic field exhibits a strong toroidal component due to the Brown Dwarf's rotation.
    \item The interior structure of the Brown Dwarf is sensitive to the initial conditions, with higher rotation rates or weaker magnetic fields resulting in flatter profiles of density, temperature, and specific entropy.
\end{enumerate}

These results demonstrate that the star-like scenario can successfully produce Brown Dwarfs, reinforcing it as a viable formation mechanism. Our simulations demonstrate the role of the initial core mass, magnetic fields, and angular momentum content in dictating the structure and evolution of nascent Brown Dwarfs. Future studies will analyze the birth of circumstellar disks around the nascent object and study the evolution of the magnetic field. This work provides a framework for theoretical models aiming to describe the formation and evolution of these very low-mass objects.

\begin{acknowledgments}
This work has received funding from the French Agence Nationale de la Recherche (ANR) through the projects DISKBUILD (ANR-20-CE49-0006), and PROMETHEE (ANR-22-CE31-0020). We gratefully acknowledge the support from the CBPsmn (PSMN, Pôle Scientifique de Modélisation Numérique) at ENS de Lyon for providing computing resources to carry-out and analyze our simulations. Simulations were also partially produced using GENCI allocations (grant A0180416201).
\end{acknowledgments}

\appendix

\section{First core outflow}
\label{appendix:outflow}

\begin{figure*}%[h]
\centering
\includegraphics[scale=.2]{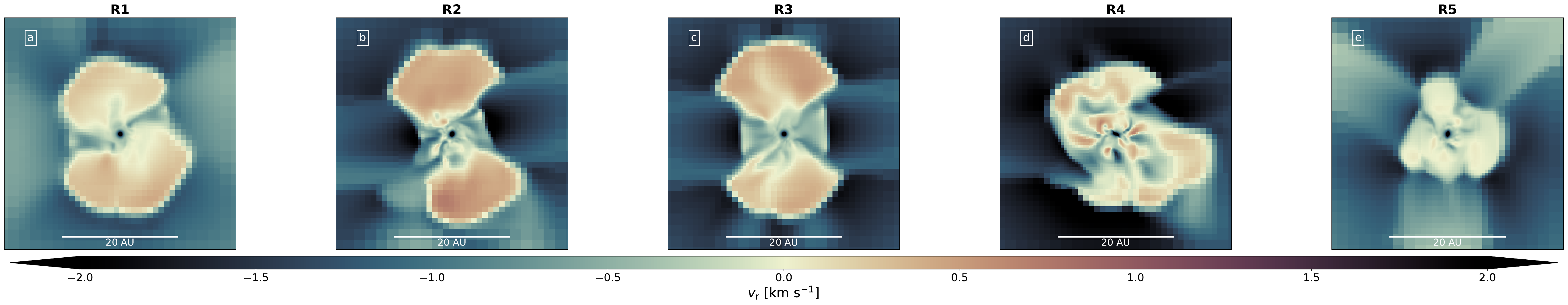} %pour fixer l'échelle.
\caption{Bipolar low velocity outflows at first core scales. Shown here are radial velocity slices for each run.}
\label{fig:outflow}
\end{figure*}

At first core scales ($\sim 20~\mathrm{AU}$), a bipolar low velocity outflow ($\sim 2~\mathrm{km\ s^{-1}}$) is launched through the magneto-centrifugal mechanism \citep{blandford_1982}. In Fig. \ref{fig:outflow}, we display radial velocity slices that clearly illustrate this bipolar outflow for runs R1-3, although it is less clearly defined for run R4 due to its weaker magnetic field strength. Finally, run R5 seems to exhibit the weakest outflow, which is due to its weaker magnetic field strength owing to the added thermal pressure support reducing the density of the first core's surroundings. These outflows are a key signature of a star-like scenario for BD formation, and expulsion of material in the form of jets and outflows have been observed around young BDs \citep{whelan_2005, phan_bao_2008, riaz_2017, whelan_2018, riaz_2019}.

\section{Resolution study}
\label{appendix:resstudy}
\begin{figure}%[h]
\centering
\includegraphics[scale=.28]{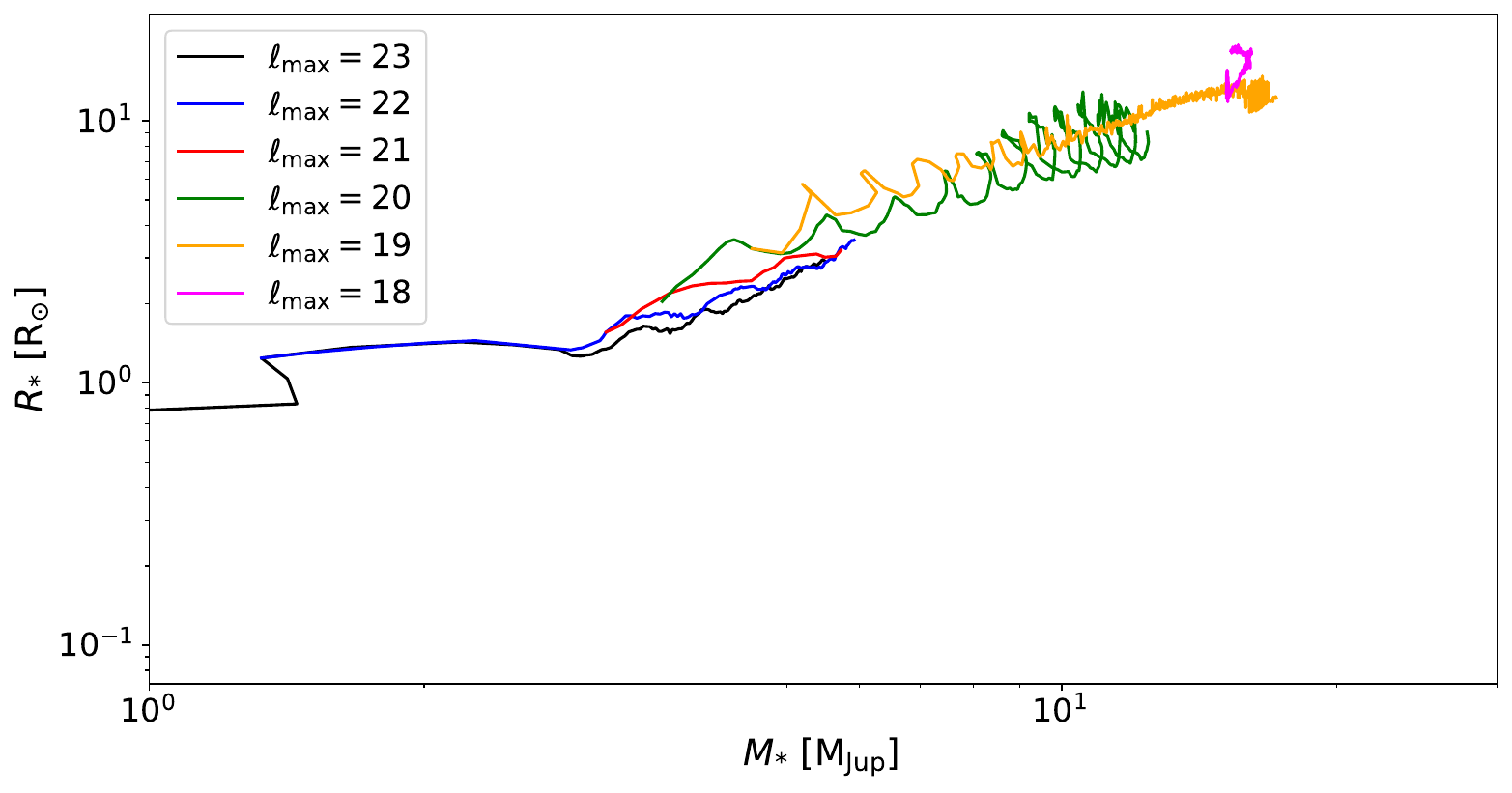} %pour fixer l'échelle.
\caption{Brown Dwarf radius as a function of mass for run R1 at different maximum refinement level. The lower resolution runs are restarts at different snapshots of their higher resolution counterparts.}
\label{fig:resstudy}
\end{figure}

To assess our simulation's convergence, we conducted a resolution study by restarting run R1 at lower resolutions from various snapshots. Figure \ref{fig:resstudy} shows the mass-radius relationship of the brown dwarf across different resolutions. Lower resolution runs systematically produce larger radii at a given mass, a consequence of reduced turbulent energy transport. This result is consistent with the findings of \cite{ahmad_2023}.

\section{Disk formation}
\label{appendix:diskform}

\begin{figure}
     \centering
     \begin{subfigure}{0.4\textwidth}
         \centering
         \includegraphics[width=\linewidth]{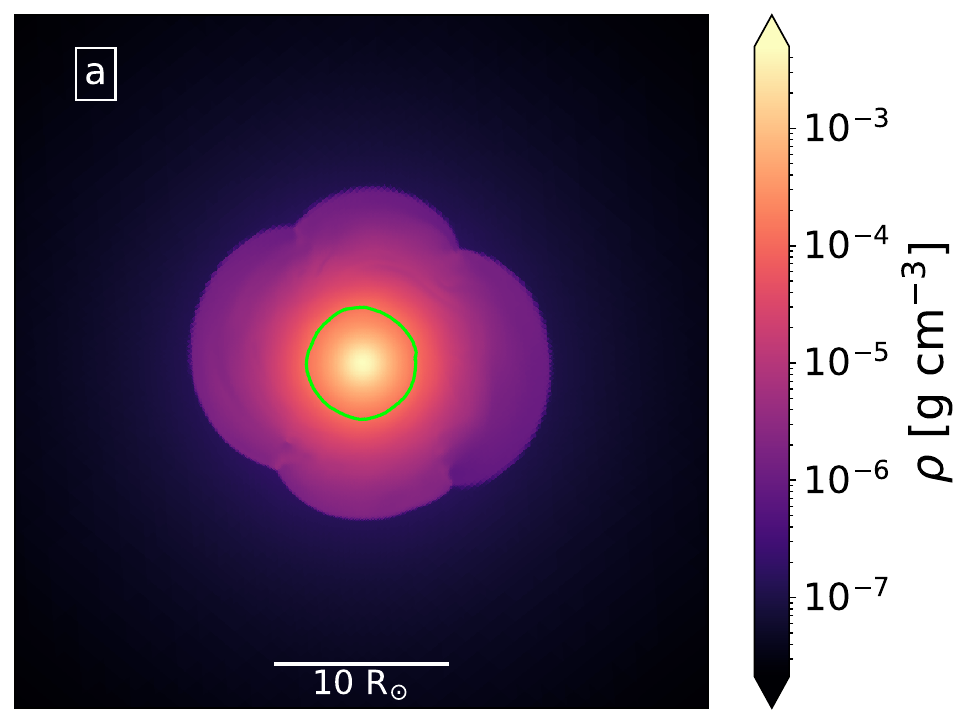}
         % \caption{Density slice }
     \end{subfigure}
     \begin{subfigure}{0.4\textwidth}
         \centering
         \includegraphics[width=\linewidth]{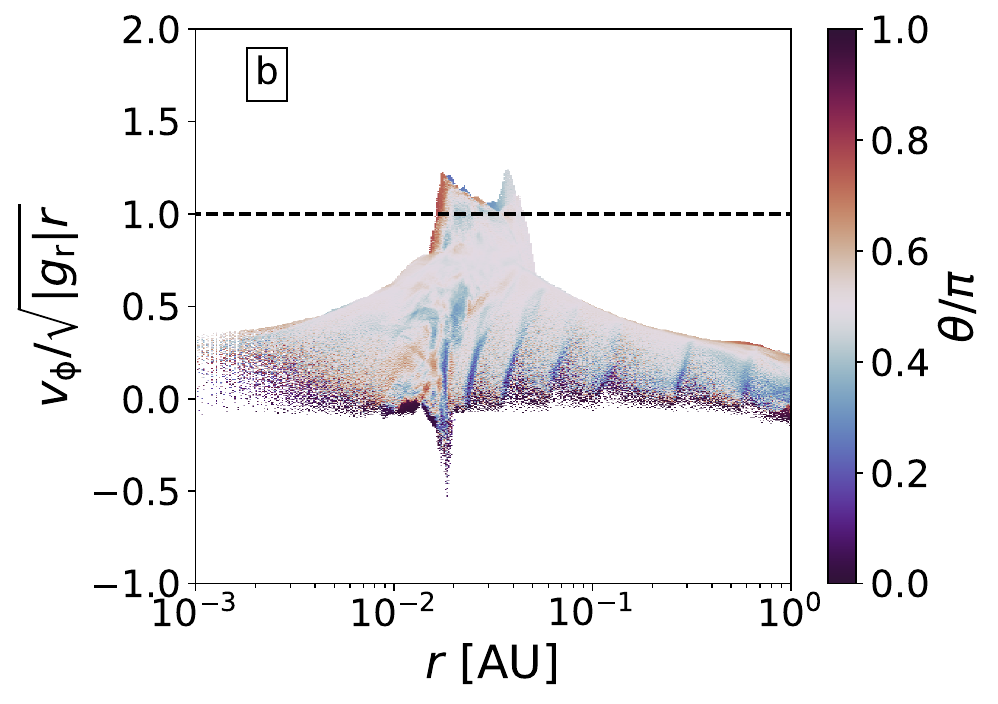}
         % \caption{histogram}
     \end{subfigure}
     \caption{Circumstellar disk formed in run R2, visualized at our final simulation snapshot. Left: Density slice across the center of the BD, visualized in a top-down view. The lime colored contour displays the BD surface, defined as all gas having dissociated over 95\% of its H$_2$. Right: 2D histogram displaying the distribution of azimuthal velocity normalized by $\sqrt{g_{\mathrm{r}}r}$ as a function of radius, binned across all cells in the computational domain. The colormap indicates the co-latitude $\theta$ scaled by $\pi$, with $\theta/\pi = 0.5$ representing the equator, while $\theta/\pi = 1$ and 0 correspond to the south and north poles, respectively. A dashed line marks the threshold $v_{\phi}/\sqrt{g_{\mathrm{r}}r}=1$.}
     \label{fig:diskform}
\end{figure}

The star-like formation scenario of brown dwarfs predicts the presence of circumstellar disks, which have been observed around BDs and planetary-mass objects \citep{pascucci_2003, luhman_2005, luhman_2016, seo_2025}. As shown in Fig. \ref{fig:BDprops}c, the BDs accrete substantial angular momentum from their surroundings immediately after birth, potentially leading to the rotational breakup of the proto-BD \citep{ahmad_2024}. This outcome is realized in our most evolved simulation (run R2). Figure \ref{fig:diskform}a displays the resulting disk at the final snapshot of run R2, where a density slice reveals a prominent rotating structure around the central BD (lime contour). This disk forms as the BD's outer layers expand outward due to excess angular momentum, as fluid parcels exceeding breakup velocity can be seen in Fig. \ref{fig:diskform}b.

\bibliography{biblio}{}
\bibliographystyle{aasjournalv7}

\end{document}